\documentclass[final,1p,times]{elsarticle}
\usepackage{graphicx}
\usepackage{amssymb}
\usepackage{amsthm}

\newcommand{\beq}{\begin{eqnarray}}
\newcommand{\eeq}{\end{eqnarray}}

\begin{document}

\begin{frontmatter}

\title{Exact sum rules for inhomogeneous systems containing a zero mode}
\author{Paolo Amore}
\ead{paolo.amore@gmail.com}
\address{Facultad de Ciencias, CUICBAS, Universidad de Colima,\\
Bernal D\'{i}az del Castillo 340, Colima, Colima, Mexico}

\begin{abstract}
We show that the formulas for the sum rules for the eigenvalues of inhomogeneous 
systems that we have obtained in two recent papers are incomplete when the system
contains a zero mode. We prove that there are finite contributions of the zero
mode to the sum rules and we explicitly calculate the expressions for the sum rules
of order one and two. The previous results for systems that do not contain a zero mode
are unaffected.
\end{abstract}
%\pacs{02.30.Mv,03.70.+k,11.10.Gh}


\end{frontmatter}
%\pacs{03.65.Ge,02.70.Jn,11.15.Tk}

\section{Introduction}

In two recent papers, refs.~\cite{Amore13a,Amore13b}, we have derived explicit expressions
for the sum rules involving the eigenvalues of inhomogeneous systems 
described by the Helmholtz equation in a finite region $\Omega$ in $d$ dimensions
\beq
(-\Delta) \Psi_n(x_1,\dots, x_d) = E_n \Sigma(x_1,\dots, x_d) \Psi_n(x_1,\dots, x_d) 
\label{Helmholtz1}
\eeq
where $\Sigma(x_1,\dots, x_d) > 0$ for $(x_1,\dots, x_d)\in \Omega$. The
eigenfunctions $\Psi_n(x_1,\dots, x_d)$ obey specific boundary conditions
on $\partial\Omega$.

As we have discussed in our previous papers Eq.~(\ref{Helmholtz1})
is isospectral to the equation
\beq
\left[ \frac{1}{\sqrt{\Sigma(x_1,\dots, x_d) }} (-\Delta ) \frac{1}{\sqrt{\Sigma(x_1,\dots, x_d) }}\right] 
\Phi_n(x_1,\dots, x_d)  = E_n \Phi_n(x_1,\dots, x_d)  \ ,
\label{Helmholtz2}
\eeq
while their eigenfunctions are simply related by 
$\Psi_n(x_1,\dots, x_d) = \Phi_n(x_1,\dots, x_d)/\sqrt{\Sigma(x_1,\dots, x_d)} $.

The spectrum of Eqs.(\ref{Helmholtz1}) and (\ref{Helmholtz2}) is bounded from below, 
in some cases being composed by strictly positive eigenvalues while in other cases  
containing also a zero mode. For example, in the case of an inhomogeneous string, discussed
in Ref.~\cite{Amore13a} a zero mode appears when either Neumann or periodic boundary
conditions are enforced.

We briefly describe the procedure that we have devised in our previous work to evaluate the sum rules 
$Z_p=\sum_n 1/E_n^p$, with $p=p_0,p_0+1,\dots$ and $p_0$ being the smallest integer 
for which the series is convergent (in one dimension $p_0=1$). 

We first define the operator
\beq
\hat{O} \equiv \frac{1}{\sqrt{\Sigma(x_1,\dots,x_d)}} (-\Delta ) \frac{1}{\sqrt{\Sigma(x_1,\dots,x_d)}}
\eeq
appearing in Eq.~(\ref{Helmholtz2}). 

The inverse operator may be formally expressed
in terms of the Green's function of the negative Laplacian obeying the same boundary conditions
\beq
\hat{O}^{-1} f = \sqrt{\Sigma(x_1,\dots,x_d)} \int_{\Omega} G(x_1,\dots,x_d,y_1,\dots,y_d ) \sqrt{\Sigma(y_1,\dots,y_d)}
f(y_1,\dots,y_d)
\eeq
Clearly the eigenvalues of this operator are just the reciprocals of the eigenvalues of Eq.~(\ref{Helmholtz2}) (for the
moment being we are assuming that the zero mode is not present).

Using the invariance of the trace of an hermitean operator under unitary transformations we are able
to relate the spectral sum rules $Z_p=\sum_n 1/E_n^p$ to the trace of $\hat{O}^{-p}$, calculated in a 
suitable basis. We have thus obtained in \cite{Amore13a,Amore13b} explicit formulas, which in some cases
may be evaluated exactly.

Although this analysis is correct for the cases of problems with a strictly positive spectrum, 
our previous results are incomplete in the case of a spectrum containing a zero mode, because of additional
contributions that we overlooked in our previous calculation. 
We will now proceed to derive these contributions and then to test them with precise numerical calculations.

In order to avoid the presence of a zero mode we consider the modified operator
\beq
\hat{O}_\gamma \equiv \frac{1}{\sqrt{\Sigma(x_1,\dots,x_d)}} (-\Delta + \gamma) \frac{1}{\sqrt{\Sigma(x_1,\dots,x_d)}}
\eeq
where $\gamma \rightarrow 0^+$; similarly we modify Eq.~(\ref{Helmholtz1}) as
\beq
(-\Delta + \gamma) \Psi_n(x_1,\dots, x_d) = E_n \Sigma(x_1,\dots, x_d) \Psi_n(x_1,\dots, x_d) 
\label{Helmholtz1b}
\eeq

The trace of $\hat{O}_\gamma$ is an invariant under unitary transformations and it may then be evaluated using 
a suitable basis, as done in the case of problem without a zero mode. Formally the traces of  order $p$, with 
$p=1,2,\dots$ may be evaluated using the same formulas of Ref.\cite{Amore13a,Amore13b}, but written in terms
of the Green's function of the shifted negative Laplacian $(-\Delta+\gamma)$
\beq
G_\gamma(x_1,\dots,x_d,y_1,\dots,y_d ) = \frac{1}{V_\Omega \gamma} + \sum_n \frac{\phi_n(x_1,\dots,x_d) \phi_n(y_1,\dots,y_d)}{\epsilon_n+\gamma}
\eeq
where  $V_\Omega$ is the volume of the region $\Omega$. 
Here $\epsilon_n$ and $\phi_n$ are the eigenvalues and eigenfunctions of the negative Laplacian on $\Omega$ 
obeying specific boundary conditions. 

It is possible to expand the Green's function around $\gamma=0$ as
\beq
G_\gamma(x_1,\dots,x_d,y_1,\dots,y_d ) = \frac{1}{V_\Omega \gamma} + 
\sum_{q=0}^\infty  (-1)^q \gamma^q G^{(q)}(x,y)
\eeq
where
\beq
G^{(q)}(x,y) \equiv \sum_{n\neq 0} \frac{\phi_n(x_1,\dots,x_d) \phi_n(y_1,\dots,y_d)}{\epsilon_n^{q+1}} 
\label{gq}
\eeq
Observe that $G^{(0)}(x,y)$ is the regularized Green's function discussed in \cite{Amore13a,Amore13b}.
 
One can easily see that, for $q=1,2,\dots$,
\beq
(- \Delta_x) G^{(q)}(x_1,\dots,x_d,y_1,\dots,y_d ) = G^{(q-1)}(x_1,\dots,x_d,y_1,\dots,y_d )
\eeq
and, for $q=0$,
\beq
(- \Delta_x) G^{(0)}(x_1,\dots,x_d,y_1,\dots,y_d ) = \delta(x_1-y_1) \dots \delta(x_d-y_d) - \frac{1}{V}
\eeq

Using these relations we find
\beq
G^{(q+1)}(x_1,\dots,x_d,y_1,\dots,y_d ) \equiv \int d^dz \ G^{(0)}(x_1,\dots,x_d,z_1,\dots,z_d ) \ G^{(q)}(z_1,\dots,z_d,y_1,\dots,y_d ) 
\eeq
which can be straightforwardly verified using the definition (\ref{gq}).

For a finite $\gamma$ these traces provide the sum rules
\beq
{\rm Tr}(\hat{O}^{-1}_\gamma)^p = \sum_{n=0}^\infty E_n^{-p}
\eeq
which include the contribution of the zero mode, while the sum rules considered in Refs.~\cite{Amore13a,Amore13b} 
are defined as $\lim_{\gamma\rightarrow 0} \sum_{n=1}^\infty E_n^{-p}$. We will now show that it is still 
possible to extract the physical sum rules by properly handling the finite contributions to the trace stemming 
from the zero mode.

Formally we may write
\beq
\sum_{n=1}^\infty E_n^{-p} = \lim_{\gamma\rightarrow 0} \frac{1}{p!}\frac{\partial^p}{\partial \gamma^p} 
\left[ \gamma^p \left( {\rm Tr}(\hat{O}^{-1}_\gamma)^p - \frac{1}{E^p_0(\gamma)}\right)\right] 
\label{corrected}
\eeq
where $E_0(\gamma)$ is the energy of the lowest mode, which depends on $\gamma$; clearly for $\gamma = 0$ 
the eigenfunction of the fundamental mode of Eq.~(\ref{Helmholtz1b}) is $\Psi_0 = constant$
and therefore $E_0(0)=0$. For an infinitesimal $\gamma$ we expect that both the eigenfunctions and eigenvalues
of Eq.~(\ref{Helmholtz1b}) have the perturbative expansions $\Psi_n(x_1,\dots, x_d) = \sum_r \gamma^r \Psi_n^{(r)}(x_1,\dots, x_d)$
and $E_n = \sum_r \gamma^r E_n^{(r)}$ respectively. 

Physically Eq.~(\ref{corrected}) takes into account the contributions due to the zero mode which are finite for $\gamma \rightarrow 0$.
These contributions are two kinds: the first term in the rhs of the equation contains the contributions due to the finite modification 
of the trace for $\gamma \rightarrow 0$, while the second term takes care of eliminating the finite contribution
due to $1/E^p_0(\gamma)$ for $\gamma \rightarrow 0$. Notice that the divergent contributions due to the zero mode are automatically 
eliminated.

Let us discuss explicitly the simplest case of the sum rule of order $p=1$ for an inhomogeneous string.
In this case
\beq
{\rm Tr}(\hat{O}^{-1}_\gamma) = {\rm Tr} \left[ \left(\frac{1}{V_\Omega \gamma} + \sum_{q=0}^\infty  (-1)^q \gamma^q G^{(q)}
\right) \Sigma\right]
\eeq

The finite part of this expression is just
\beq
\lim_{\gamma\rightarrow 0} \frac{\partial}{\partial \gamma}  \gamma {\rm Tr}(\hat{O}^{-1}_\gamma) = 
{\rm Tr} \left[ G^{(0)} \Sigma\right]
\eeq
and it coincides with the expression in Ref.~\cite{Amore13a}. However in order to obtain the sum rule we also need to subtract 
the finite contribution of the zero mode. To do this we need to evaluate $E_0(\gamma)$ to order $\gamma^2$ and use it to obtain
\beq
\frac{1}{E_0(\gamma)} = \frac{1}{E_0^{(1)} \gamma + E_0^{(2)} \gamma^2 + \dots} = \frac{1}{E_0^{(1)} \gamma} -
\frac{E_0^{(2)}}{\left(E_0^{(1)}\right)^2} + O(\gamma)
\eeq

Using the perturbative approach described in \ref{appA} we obtain:
\beq
\frac{1}{E_0} &=& \frac{\left(\int_\Omega \Sigma(x_1,\dots, x_d) d^dx\right)}{V_\Omega \gamma} \nonumber \\
&+& 
\frac{\int_\Omega d^dx \int_\Omega d^dy \Sigma(x_1,\dots, x_d) G^{(0)}(x_1,\dots,x_d,y_1,\dots,y_d ) \Sigma(y_1,\dots, y_d)}{\left(\int_\Omega \Sigma(x_1,\dots, x_d) d^dx\right)} + O\left( \gamma\right)
\eeq

The finite contribution of the zero mode is therefore
\beq
\left.\frac{1}{E_0}\right|_{finite} &=& 
\left(\frac{\int_\Omega d^dx \int_\Omega d^dy \Sigma(x_1,\dots, x_d) G^{(0)}(x_1,\dots,x_d,y_1,\dots,y_d ) \Sigma(y_1,\dots, y_d)}{\left(\int_\Omega \Sigma(x_1,\dots, x_d) d^dx\right)} \right)
\eeq
and it needs to be subtracted from the expressions for $Z(1)$ obtained in Ref.\cite{Amore13a}
for Neumann and periodic bc; in the case of an inhomogeneous string with Neumann bc one has
\beq
Z^{(NN)}_1 = \int_{-a/2}^{a/2} \Sigma(x) \left(\frac{a}{12}+\frac{x^2}{a}\right) dx -
\frac{\int_{-a/2}^{a/2}  dx \int_{-a/2}^{a/2}  dy \Sigma(x) G_{NN}^{(0)}(x,y ) \Sigma(y)}{\int_{-a/2}^{a/2}  dx \Sigma(x)} 
\label{znn1}
\eeq
while, for the case of periodic bc one has
\beq
Z^{(PP)}_1 = \int_{-a/2}^{a/2} \frac{a}{12} \Sigma(x)  dx -
\frac{\int_{-a/2}^{a/2}  dx \int_{-a/2}^{a/2}  dy \Sigma(x) G_{PP}^{(0)}(x,y ) \Sigma(y)}{\int_{-a/2}^{a/2}  dx \Sigma(x)} 
\label{zpp1}
\eeq

Let us now discuss the sum rule of order two; we need the trace
\beq
{\rm Tr}(\hat{O}^{-2}_\gamma) = {\rm Tr} \left[ \left(\frac{1}{V_\Omega \gamma} + \sum_{q=0}^\infty  (-1)^q \gamma^q G^{(q)}
\right) \Sigma \left(\frac{1}{V_\Omega \gamma} + \sum_{q=0}^\infty  (-1)^q \gamma^q G^{(q)}
\right) \Sigma \right]
\eeq
and we extract the finite part of this expression with the limit
\beq
\lim_{\gamma\rightarrow 0} \frac{1}{2}\frac{\partial^2}{\partial \gamma^2}  \gamma^2 {\rm Tr}(\hat{O}^{-2}_\gamma) = 
{\rm Tr} \left[ G^{(0)} \Sigma G^{(0)} \Sigma\right] - \frac{2}{V_\Omega}  {\rm Tr} \left[ \Sigma G^{(1)} \Sigma\right] 
\eeq

Notice that the first term in the rhs of the equation is the term obtained in Refs.~\cite{Amore13a,Amore13b}: the second term is due 
to a finite contribution of the zero mode to the eigenvalues and it involves the Green's function $G^{(1)}$.

We have obtained the explicit expressions for $G^{(1)}(x,y)$ for Neumann and periodic bc in one dimension,
which read
\beq
G_{NN}^{(1)}(x,y) &=& \left\{\begin{array}{ccc}
\frac{a^4-30 a^2 \left(x^2-6 x y+y^2\right)-60 a (x-y)^3-30 \left(x^4+6 x^2 y^2+y^4\right)}{720 a} & , & x< y \\
\frac{a^4-30 a^2 \left(x^2-6 x y+y^2\right)+60 a (x-y)^3-30 \left(x^4+6 x^2 y^2+y^4\right)}{720 a} & , & x>y \\
\end{array}
\right.
\eeq
and
\beq
G_{PP}^{(1)}(x,y) &=& \left\{\begin{array}{ccc}
\frac{a^4-30 a^2 (x-y)^2-60 a (x-y)^3-30 (x-y)^4}{720 a} & , & x< y \\
\frac{a^4-30 a^2 (x-y)^2+60 a (x-y)^3-30 (x-y)^4}{720 a} & , & x>y \\
\end{array}
\right.
\eeq
where $|x|\leq a/2$ and $|y|\leq a/2$.

In \ref{appA} we calculate explicitly the expression for the energy of the zero mode up to third order
and we may use it in the calculation of the sum rule of order $2$ isolating the term independent of
$\gamma$ in $E_0^{-2}$:
\beq
E_0^{-2} = \frac{1}{\left(\gamma E_0^{(1)}\right)^2} - \frac{2 E_0^{(2)}}{\gamma\left( E_0^{(1)}\right)^3} +
\frac{ 3 \left( E_0^{(2)}\right)^2 - 2 E_0^{(1)} E_0^{(3)} }{\left( E_0^{(1)}\right)^4} + \dots
\eeq
The explicit expressions for $E_0^{(1)}$, $E_0^{(2)}$ and $E_0^{(3)}$ are given in \ref{appA}.

We are now in position of writing the sum rule of order 2 as
\beq
\sum_{n=1}^\infty \frac{1}{E_n^2} = 
{\rm Tr} \left[ G^{(0)} \Sigma G^{(0)} \Sigma\right] - \frac{2}{V_\Omega}  {\rm Tr} \left[ \Sigma G^{(1)} \Sigma\right] 
 - \frac{ 3 \left( E_0^{(2)}\right)^2 - 2 E_0^{(1)} E_0^{(3)} }{\left( E_0^{(1)}\right)^4} 
\eeq
where only the first term of the rhs of the equation was considered in our previous work.

Clearly the calculation of the sum rules of higher order can be carried out 
using the general procedure that we have described.

%%%%%%%%%%%%%%%%%%%%%%%%%%%%%%%%%%%%%%%%%%%%%%%%%%%%%%%%%%%%%%%%%%%%%%%%%%%%%%%%%%%%%%%%%%%%%%%%%%%%%%%%%%%%%%

\section{Applications}

We review three of the examples previously discussed in Ref.\cite{Amore13a,Amore13b} and calculate the finite
contribution of the zero mode.

\subsection{Isospectral strings}

The first example studied in Ref.\cite{Amore13a} was the string with density
\beq
\Sigma(x) = \frac{(1+\alpha)^2}{(1+\alpha (x+1/2))^4} \ \ \ , \ \ \ |x| \leq 1/2
\eeq
which for Dirichlet boundary conditions is known as the "Borg string" and it is 
isospectral to the uniform string.

In light of our previous discussion, we need to modify Eqs.(38) and (41) of that paper
using Eqs.(\ref{znn1}) and (\ref{zpp1}); a simple calculation provides
\beq
Z_1^{(NN)}(\alpha) &=& \frac{\alpha ^2+5 \alpha +5}{10 \left(\alpha ^2+3 \alpha +3\right)} \\
Z_1^{(PP)}(\alpha) &=& \frac{5 a (\alpha  (\alpha +3)+3)^2-\alpha ^2 (\alpha  (5
   \alpha +12)+12)}{180 (\alpha +1) (\alpha^2 + 3 \alpha+3)}
\eeq

Notice that these sum rules are still dependent on $\alpha$, contrary to the case of Dirichlet
boundary conditions, which is isospectral to the uniform string.

It is possible to test numerically these results with precision, given that the eigenvalues
in both cases are solutions of transcendental equations and therefore one can calculate 
accurately a large number of them~\cite{Gottlieb89,Merzbacher70}.

We have calculated the first $10000$ numerical Neumann eigenvalues for $\alpha=1$ with $200$ 
digits of accuracy, and we have used the last $100$ of them to estimate the 
asymptotic behavior $E_n \approx c_1 n^2 + c_2 + c_3/n^2+c_4/n^4$. The spectral sum rule
is then estimated numerically
\beq
Z_1^{(NN)}(1) &\approx& \sum_{n=1}^{10000} \frac{1}{E_n^{(num)}} + \sum_{n=10001}^\infty \frac{1}{c_1 n^2 + c_2 + c_3/n^2+c_4/n^4} \nonumber \\
&=& 0.157142857142857142857142857142857142857127
\eeq 

The exact result obtained from Eq.(\ref{znn1}) with $\alpha=1$ is
\beq
Z_1^{(NN)}(1) = \frac{11}{70} \approx 0.157142857142857142857142857142857142857143
\eeq

In the case of periodic bc, we have calculated numerically the first $20000$ eigenvalues, with the same accuracy as before.
We have also estimated the asymptotic behavior $E_n \approx c_1 n^2 + c_2 + \sum_{p=1}^5 c_{p+2}/n^{2p}$
using the last $10000$ eigenvalues.

The spectral sum rule is then estimated numerically
\beq
Z_1^{(PP)}(1) &\approx& \sum_{n=1}^{20000} \frac{1}{E_n^{(num)}} + \sum_{n=20001}^\infty \frac{1}{c_1 n^2 + c_2 + \sum_{p=1}^5 c_{p+2}/n^{2p}} \nonumber \\
&=& 0.085714285714961 
\eeq 

The exact result obtained from Eq.(\ref{zpp1}) with $\alpha=1$ is
\beq
Z_1^{(PP)}(1) = \frac{3}{35} \approx 0.085714285714286
\eeq

The convergence of the numerical estimate towards the exact value is slower in the case of periodic bc.

We have also calculated the sum rules of order 2 obtaining 
\beq
Z_2^{(NN)}(\alpha) = \frac{\alpha^4+10 \alpha^3+45 \alpha^2+70 \alpha+35}{350 \left(\alpha^2+3 \alpha+3\right)^2}
\eeq

We have estimated  numerically the sum rule as in the previous case obtaining
\beq
Z_2^{(NN)}(1) &\approx& \sum_{n=1}^{10000} \frac{1}{(E_n^{(num)})^2} + \sum_{n=10001}^\infty \frac{1}{(c_1 n^2 + c_2 + c_3/n^2+c_4/n^4)^2} 
\nonumber \\
&=& 0.009387755102040816326530612244897959183673469387752189
\eeq 
which can be compared with the exact value
\beq
Z_2^{(NN)}(1) &=& \frac{23}{2450} \nonumber\\
&\approx& 0.009387755102040816326530612244897959183673469387755102
\eeq

In the case of periodic bc we have obtained
\beq
Z_2^{(PP)}(\alpha) =\frac{24 \alpha^4+100 \alpha^3+205 \alpha^2+210 \alpha+105}{8400 \left(\alpha^2+3 \alpha+3\right)^2}
\eeq

The spectral sum rule is then estimated numerically
\beq
Z_2^{(PP)}(1) &\approx& \sum_{n=1}^{20000} \frac{1}{(E_n^{(num)})^2} + \sum_{n=20001}^\infty 
\frac{1}{(c_1 n^2 + c_2 + \sum_{p=1}^5 c_{p+2}/n^{2p})^2} \nonumber \\
&=& 0.0015646258503401360550515
\eeq 
which can be compared to the exact value
\beq
Z_2^{(PP)}(1) &=& \frac{23}{14700} \approx 0.0015646258503401360544218
\eeq

%%%%%%%%%%%%%%%%%%%%%%%%%%%%%%%%%%%%%%%%%%%%%%%%%%%%%%%%%%%%%%%%%%%%%%%%%%%%%%%%%%%%%%%
\subsection{A string with rapidly oscillating density}

In Ref.\cite{Amore13a} we have considered a string with a rapidly oscillating density
\beq
\Sigma(x) = 2 + \sin \left( 2 + \frac{2\pi (x+1/2)}{\varepsilon}\right)
\eeq
with $\varepsilon \rightarrow 0^+$ and $|x|\leq 1/2$. 

We need to revise our results for the case of Neumann bc in light of 
the discussion of the contributions of the zero mode done in the present 
paper.

In this case we find
\beq
E_0^{(NN)} = E_0^{(1NN)} \gamma + E_0^{(2NN)} \gamma^2 + E_0^{(3NN)} \gamma^3 + \dots
\eeq
where
\beq
E_0^{(1NN)}  &=& \frac{\pi }{\varepsilon \sin ^2\left(\frac{\pi}{\varepsilon}\right)+2 \pi } \\
E_0^{(2NN)}  &=& \frac{\varepsilon^2 \left(18
   \varepsilon^2-8 \left(3
   \varepsilon^2+\pi ^2\right) \cos \left(\frac{2
   \pi }{\varepsilon}\right)+\left(6
   \varepsilon^2-4 \pi ^2\right) \cos
   \left(\frac{4 \pi }{\varepsilon}\right)+9 \pi 
   \varepsilon \sin \left(\frac{4 \pi
   }{\varepsilon}\right)-24 \pi ^2\right)}{96 \pi
    \left(\varepsilon \sin ^2\left(\frac{\pi
   }{\varepsilon}\right)+2 \pi \right)^3} \\
E_0^{(3NN)}  &=& \frac{1}{5760 \pi ^3 \left(\varepsilon+\varepsilon \left(-\cos \left(\frac{2 \pi}{\varepsilon}\right)\right)+4 \pi \right)^5}
\left[ \varepsilon^3 \left(45 \pi \left(\varepsilon^2-16 \pi ^2\right) \varepsilon^2 \sin \left(\frac{8 \pi }{\varepsilon}\right) \right. \right.\nonumber \\
&-& \left. \left. 24 \left(15 \varepsilon^4+35 \pi ^2 \varepsilon^2-8 \pi ^4\right) \varepsilon \cos \left(\frac{8 \pi }{\varepsilon}\right)-180 \pi 
 \left(\varepsilon^3-94 \pi \varepsilon^2+20 \pi ^2 \varepsilon+32 \pi ^3\right) \varepsilon \sin \left(\frac{2\pi }{\varepsilon}\right)\right. \right.\nonumber \\
&-& \left. \left. 90 \pi  \left(-3 \varepsilon^3+376 \pi  \varepsilon^2+192 \pi ^2 \varepsilon+64 \pi ^3\right)
   \varepsilon \sin \left(\frac{4 \pi}{\varepsilon}\right)-180 \pi \left(\varepsilon^3-94 \pi  \varepsilon^2+20 \pi ^2 \varepsilon+32 \pi ^3\right) \varepsilon \sin \left(\frac{6\pi }{\varepsilon}\right)\right. \right.\nonumber \\
&+& \left. \left. 24 \left(-420 \varepsilon^5-2160 \pi  \varepsilon^4-95 \pi ^2 \varepsilon^3+384 \pi ^4 \varepsilon+64 \pi ^5\right) \cos\left(\frac{4 \pi }{\varepsilon}\right) \right. \right.\nonumber \\
&+& \left. \left. 24 \left(840 \varepsilon^4+5400 \pi  \varepsilon^3+445 \pi ^2 \varepsilon^2-1860 \pi ^3 \varepsilon+712 \pi ^4\right) \varepsilon \cos \left(\frac{2 \pi}{\varepsilon}\right) \right. \right.\nonumber \\
&-& \left. \left. 40 \left(315 \varepsilon^5+2160 \pi \varepsilon^4+150 \pi ^2 \varepsilon^3-1464 \pi ^3 \varepsilon^2-600 \pi ^4 \varepsilon+64 \pi ^5\right) \right. \right.\nonumber \\
&+& \left. \left. 8 \left(360 \varepsilon^5+1080 \pi  \varepsilon^4-195 \pi ^2 \varepsilon^3-1740 \pi ^3\varepsilon^2+168 \pi ^4 \varepsilon+128 \pi ^5\right) \cos \left(\frac{6 \pi}{\varepsilon}\right)\right)
\right]
\eeq

It is important to notice that the perturbative corrections $E_0^{(qNN)}$  obey a the hierarchy $E_0^{(1NN)}  \gg E_0^{(2NN)} \gg E_0^{(3NN)} \gg \dots$
for $\varepsilon\rightarrow 0$, since
\beq
E_0^{(1NN)}  &\approx&  \frac{1}{2}-\frac{\varepsilon \sin
   ^2\left(\frac{\pi }{\varepsilon}\right)}{4 \pi} + \dots \\
E_0^{(2NN)}  &\approx& -\frac{\varepsilon^2 \left(2 \cos \left(\frac{2
   \pi }{\varepsilon}\right)+\cos \left(\frac{4
   \pi }{\varepsilon}\right)+6\right)}{192 \pi
   ^2} + \dots \\
E_0^{(3NN)}  &\approx&    \frac{\varepsilon^3 \left(3 \cos \left(\frac{4
   \pi }{\varepsilon}\right)+2 \cos \left(\frac{6
   \pi }{\varepsilon}\right)-5\right)}{11520 \pi
   ^3} + \dots 
\eeq

Because of this behavior, the corrections to the sum rule for the Neumann eigenvalues that we have discussed in this paper are 
negligible for $\varepsilon\rightarrow 0$, and in this limit the general formulas of Ref.\cite{Amore13a} should dominate.
Unfortunately, an error affected Eq.(53) of Ref.\cite{Amore13a}, and therefore Fig.6. 

We have been able to calculate explicitly the first two sum rules for this string; in particular 
the sum rule of order $1$ reads
\beq
Z_1^{(NN)}(\varepsilon) &=& \frac{\varepsilon \sin \left(\frac{2 \pi
   }{\varepsilon}\right) \left(\left(2 \pi ^2-3
   \varepsilon^2\right) \sin \left(\frac{\pi
   }{\varepsilon}\right)+3 \pi  \varepsilon
   \cos \left(\frac{\pi
   }{\varepsilon}\right)\right)+2 \pi ^3}{6 \pi
   ^3}\nonumber\\
   &+& \frac{\varepsilon^2 \left(18
   \varepsilon^2-8 \left(3
   \varepsilon^2+\pi ^2\right) \cos \left(\frac{2
   \pi }{\varepsilon}\right)+\left(6
   \varepsilon^2-4 \pi ^2\right) \cos
   \left(\frac{4 \pi }{\varepsilon}\right)+9 \pi 
   \varepsilon \sin \left(\frac{4 \pi
   }{\varepsilon}\right)-24 \pi ^2\right)}{96 \pi
   ^3 \left(\varepsilon \sin ^2\left(\frac{\pi
   }{\varepsilon}\right)+2 \pi \right)}
\eeq

Although we also dispose of an analytic expression for the sum rule of order $2$, it is
quite complicated and we do not report it here; we rather report the leading behavior of
the sum rule for $\varepsilon\rightarrow 0$, which reads
\beq
Z_2^{(NN)}(\varepsilon) &\approx&
\frac{2}{45} +\frac{2 \varepsilon \left(\cos
   \left(\frac{\pi }{\varepsilon}\right)-\cos
   \left(\frac{3 \pi
   }{\varepsilon}\right)\right)}{45 \pi
   }\nonumber\\
&+& \frac{\varepsilon^2 \left(12 \cos ^2\left(\frac{2
   \pi }{\varepsilon}\right)-9 \cos \left(\frac{2
   \pi }{\varepsilon}\right)-15 \cos
   \left(\frac{4 \pi }{\varepsilon}\right)+10
   \cos \left(\frac{6 \pi
   }{\varepsilon}\right)+32\right)}{720 \pi
   ^2} +\dots
\eeq

We have also performed a numerical test of these expressions, comparing the values of the exact sum rules
at $\varepsilon = 1$ with the approximate sum rules obtained calculating the eigenvalues numerically using a
Rayleigh-Ritz approach, with $100$ states.

The numerical estimates that we obtain with the Rayleigh-Ritz method are
\beq
Z_1^{(NN)}(1) &\approx& \sum_{n=1}^{100} \frac{1}{E_n^{(RR)}} \approx 0.31229
\eeq
and
\beq
Z_2^{(NN)}(1) &\approx&  \sum_{n=1}^{100} \frac{1}{\left( E_n^{(RR)}\right)^2} \approx 0.037798617 
\eeq

These results must be compared with the exact values
\beq
Z_1^{(NN)}(1) &=& \frac{1}{3}-\frac{3}{16 \pi ^2}  \approx 0.31433561140039500119
\eeq
and
\beq
Z_2^{(NN)}(1) &=& \frac{2}{45}-\frac{271}{256 \pi ^4}+\frac{1}{24 \pi ^2} \approx 0.037798655777122106134
\eeq
%%%%%%%%%%%%%%%%%%%%%%%%%%%%%%%%%%%%%%%%%%%%%%%%%%%%%%%%%%%%%%%%%%%%%%%%%%%%%%%%%%%%%%%

\subsection{Circular annulus}

In light of the results obtained in the present paper we need to discuss the sum rule for a circular 
annulus with Neumann boundary conditions at the border that we have recently obtained in Ref.\cite{Amore13b}.

As we have seen before, the an annulus of radii $r_{min}$ and $r_{max}=1$ may be mapped conformally to a rectangle 
of sides $a=\log 1/r_{min}$ and $b=2\pi$ by the map $f(z) = e^{z+\frac{1}{2} \log r_{min}}$. We define
$V= ab = 2 \pi \log \frac{1}{r_{min}}$ to be the area of the rectangle.
In this case one needs to solve the Helmholtz equation with the  nonhomogeneous density
$\Sigma(x,y) = r_{min} e^{2x}$.

In our calculation we will use the basis of the negative Laplacian on the rectangle, represented 
by the eigenfunctions
\beq
\Psi_{n_x,u_x,n_y,u_y}(x,y) = \psi_{n_x,u_x}^{(N)}(x) \ \phi_{n_y,u_y}^{(P)}(y)
\eeq
where
\beq
\psi_{n_x,u_x}^{(N)}(x) &=& \left\{\begin{array}{ccc}
\sqrt{\frac{1}{a}} & , & n_x = 0 \ , \ u_x=1 \\
\sqrt{\frac{2}{a}} \cos \frac{2 n_x \pi x}{a} & , & n_x>0 \ , \ u_x=1 \\
\sqrt{\frac{2}{a}} \sin \frac{(2 n_x-1) \pi x}{a} & , & n_x \geq 0 \ , \ u_x=2 \\
\end{array}
\right. \\
\phi_{n_y,u_y}^{(P)}(y) &=& \left\{\begin{array}{ccc}
\sqrt{\frac{1}{b}} & , & n_y = 0 \ , \ u_y=1 \\
\sqrt{\frac{2}{b}} \cos \frac{2 n_y \pi y}{b} & , & n_y>0 \ , \ u_y=1 \\
\sqrt{\frac{2}{b}} \sin \frac{2 n_y \pi y}{b} & , & n_y \geq 0 \ , \ u_y=2 \\
\end{array}
\right. 
\eeq

The eigenvalues of the negative Laplacian on the rectangle are
\beq
\mathcal{E}_{n_x,u_x,n_y,u_y} = \epsilon_{n_x,u_x}^{(N)} + \eta_{n_y,u_y}^{(P)} 
\eeq
where
\beq
\epsilon_{n_x,u_x}^{(N)} &=& \left\{ \begin{array}{ccccc}
\frac{4 n_x^2\pi^2}{a^2}     & , & u_x=1 & , & n_x = 0,1,2,\dots \\
\frac{(2 n_x-1)^2\pi^2}{a^2} & , & u_x=2 & , & n_x = 1,2,\dots\\
\end{array}
\right.   \\
\eta_{n_y,u_y}^{(P)} &=&  \left\{ \begin{array}{ccccc}
0   & , & u_y=1 & , & n_y = 0 \\
\frac{4 \pi^2 n_y^2}{b^2}   & , & u_y=1,2 & , & n_x = 1,2,\dots \\
\end{array}
\right.   \\
\nonumber
\eeq

It is useful to evaluate the matrix elements of the density in this basis: taking into account
the fact that $\Sigma$ does not depend on $y$, we have
\beq
\langle n_x u_x n_y u_y  | \Sigma | n'_x u'_x n'_y u'_y \rangle &=& \delta_{n_y n'_y} \delta_{u_y u'_y}
\int_{-a/2}^{a/2} dx \ \psi_{n_x,u_x}^{(N)}(x) \Sigma(x) \psi_{n'_x,u'_x}^{(N)}(x)  \nonumber 
\eeq
and 
\beq
\int_{-a/2}^{a/2} dx \ \int_{-b/2}^{b/2} dy \ \Sigma(x) \psi_{n_x,u_x}^{(N)}(x) \phi_{n_y,u_y}^{(P)}(y) = 
\sqrt{b} \delta_{n_y,0} \delta_{u_y,1} 
\langle 0 1 0 1  | \Sigma | n_x u_x 0 1\rangle
\eeq

We do not report the explicit expressions for these matrix elements, since they can be calculated easily.

We now apply the formulas of \ref{appA} to this problem; to first order we have
\beq
E_0^{(1)} = \frac{V}{\int_V dxdy  \Sigma(x)} = \frac{- \log r_{min}}{1-r_{min}^2}
\eeq
where $\int_V dxdy \equiv \int_{-a/2}^{a/2} dx \int_{-b/2}^{b/2} dy$.

To second order the energy of the zero mode becomes
\beq
E_0^{(2)} &=& - \frac{V E_0^{(1)}}{ \left(\int_V dxdy  \Sigma(x)\right)^2} \ \sum_m^\prime \frac{\langle 0 | \Sigma | m\rangle \langle m | \Sigma | 0\rangle}{\epsilon_m} \nonumber\\
&=& \frac{\log (r_{min}) \left(6 \left(r_{min}^2-1\right)^2+\log (r_{min})
   \left(-9 r_{min}^4+4 \left(r_{min}^4+r_{min}^2+1\right) \log
   (r_{min})+9\right)\right)}{6 \left(1-r_{min}^2\right)^3}
\eeq

Similarly, to third order we obtain
\beq
E_0^{(3)} &=& -\frac{\left(r_{min}^2+1\right) \log ^2(r_{min})}{16
   \left(r_{min}^2-1\right)^2}+\frac{\log (r_{min})}{2-2
   r_{min}^2}+\frac{\left(14 r_{min}^4+41 r_{min}^2+14\right) \log
   ^3(r_{min})}{12 \left(r_{min}^2-1\right)^3} \nonumber \\
&-& \frac{\left(2 r_{min}^6+7
   r_{min}^4+7 r_{min}^2+2\right) \log ^4(r_{min})}{2
   \left(r_{min}^2-1\right)^4}+\frac{2 \left(2 r_{min}^8+7 r_{min}^6+12
   r_{min}^4+7 r_{min}^2+2\right) \log ^5(r_{min})}{15
   \left(r_{min}^2-1\right)^5}
\eeq

Finally, we calculate the finite contribution of the zero mode to the trace:
\beq
-\frac{2}{V} Tr\left[ \Sigma G^{(1)} \Sigma \right] &=& -\frac{2}{V} \int_{-a/2}^{a/2} dx_1  \int_{-b/2}^{b/2} dy_1 \int_{-a/2}^{a/2} dx_2  \int_{-b/2}^{b/2} dy_2
\ \Sigma(x_1) G^{(1)}(x_1,y_1,x_1,y_1) \ \Sigma(x_2)\nonumber \\
&=&  \frac{r_{min}^4}{12}-\frac{1}{90} r_{min}^4 \log ^2(r_{min})+\frac{3
   r_{min}^4}{32 \log ^2(r_{min})}-\frac{5 r_{min}^4}{32 \log
   (r_{min})}+\frac{r_{min}^2}{12}  \nonumber\\
&-& \frac{7}{360} r_{min}^2 \log
   ^2(r_{min})-\frac{3 r_{min}^2}{16 \log ^2(r_{min})}-\frac{\log
   ^2(r_{min})}{90}+\frac{3}{32 \log ^2(r_{min})} \nonumber\\
&+& \frac{5}{32 \log
   (r_{min})}+\frac{1}{12}
\eeq   

\begin{figure}
\begin{center}
\bigskip\bigskip\bigskip
\includegraphics[width=10cm]{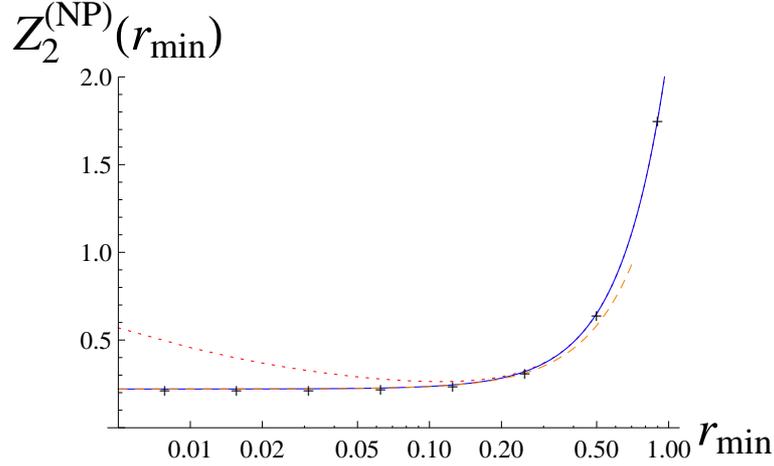}
\caption{Sum rule of order two for circular annulus with internal radius $r_{min}$ (and external radius $r_{max}=1$).
The solid blue line is the exact sum rule calculated in the present paper; the dotted red line is the result of Ref.\cite{Amore13b}, which
does not include the contributions of the zero mode; the dashed orange line is the asymptotic behavior of Eq.(\ref{asym}). The crosses 
are the precise numerical values obtained calculating a large number of eigenvalues numerically.
}
\label{Fig_1}
\end{center}
\end{figure}

Using these results  we now have an explicit expression
for the exact sum rule of order two of a circular annulus with Neumann bc at the borders. In particular,
for $r_{min}\rightarrow 0$ one has
\beq
Z_2^{(NP)}(r_{min}) \approx \frac{5 \pi ^2}{48}-\frac{155}{192} + \frac{139 r_{min}^2}{96}+ \dots
\label{asym}
\eeq
where no logarithmic divergence is present. In Fig.\ref{Fig_1} we compare the results of the present paper with 
the result of Ref.\cite{Amore13b}. Notice that the effect of the zero mode is important 
for $r_{min} \rightarrow 0$, where it cancels a divergent behavior, but completely negligible for $r_{min} \rightarrow 1$.

Incidentally we have calculated numerically the sum rule of order two for a unit circle with Neumann boundary conditions
at its border, obtaining the first $2000$ eigenvalues with an accuracy of 10 digits and estimating the contribution of the higher modes
with Weyl's law. 

We have found
\beq
\left. Z_2^{(N)}\right|_{circle} \approx 0.2207921258
\eeq
which should be compared with
\beq
Z_2^{(NP)}(0) = \frac{5 \pi ^2}{48}-\frac{155}{192} \approx 0.2207921251
\eeq
It is reasonable to conjecture that $\left. Z_2^{(N)}\right|_{circle}=\frac{5 \pi ^2}{48}-\frac{155}{192}$.

\section{Conclusions}

We have found out that the formulas for the sum rules involving eigenvalues of inhomogeneous systems 
that we have recently derived in Refs.\cite{Amore13a,Amore13b} are incomplete when the system has a zero mode. 
In one dimension this affects problems with Neumann or periodic bc, while the cases of Dirichlet or mixed 
Dirichlet-Neumann bc are unaffected.

In this paper we have derived the formalism which allows one to calculate the sum rules for systems 
containing a zero mode  exactly, by considering a nonsingular problem where the negative Laplacian 
is shifted infinitesimally and by then properly handling the contributions 
of the zero mode which are finite when the shift vanishes. Our previous approach, which used a
Green's function that did not contain the zero mode, did not account for these finite contributions.
We have explicitly obtained the formulas for the sum rules of order one and two; the calculation of
the higher order sum rules can be performed analogously and it also involves the calculation of higher
order terms in the perturbative expansion of $E_0$.

The formulas derived in this paper are compared with precise numerical results obtained for 
an exactly solvable problem, providing a useful numerical test.

%%%%%%%%%%%%%%%%%%%%%%%%%%%%%%%%%%%%%%%%%%%%%%%%%%%%%%%%%%%%%%%%%%%%%%%%%%%%%%%%%%%%%%%%%%%%%%%%%%%%%%%%%%%%%%

\appendix

\section{Perturbative calculation of $E_0$}
\label{appA}

We describe here a perturbative approach to the calculation of the eigenvalue and eigenfunction
of the lowest mode of Eq.~(\ref{Helmholtz1b}).
 
Assuming $|\gamma|\ll 1$ we write
\beq
E_0 &=& \sum_{n=0}^\infty E_0^{(n)} \gamma^n \\
\Psi_0(x_1,\dots,x_d) &=& \sum_{n=0}^\infty \Psi_0^{(n)}(x_1,\dots,x_d) \gamma^n  \nonumber 
\eeq

Let us call $\left\{ \phi_n(x_1,\dots,x_d)\right\}$ the basis obeying the boundary conditions
of the problem; it is easy to see that
\beq
E_0^{(0)} &=& 0 \\
\Psi_0^{(0)}(x_1,\dots,x_d) &=& \phi_0(x_1,\dots,x_d) = \frac{1}{\sqrt{V_\Omega}}
\eeq

To first order one obtains the equation
\beq
(-\Delta) \Psi_0^{(1)}(x_1,\dots,x_d) + \Psi_0^{(0)}(x_1,\dots,x_d)  = 
E_0^{(1)} \Sigma(x_1,\dots,x_d) \Psi_0^{(0)}(x_1,\dots,x_d) 
\eeq
which can be solved requiring 
\beq
E_0^{(1)} &=& \frac{V_\Omega}{\int_{\Omega_d} d^dx \Sigma(x_1,\dots,x_d)} \\
\Psi_0^{(1)}(x_1,\dots,x_d) &=& \sum_{m}^\prime  c_m^{(1)} \phi_m(x_1,\dots,x_d) \nonumber \\
&=&  \frac{E_0^{(1)}}{\sqrt{V_\Omega}} \sum_{m}^\prime \frac{\int_{\Omega_d} d^dy \Sigma(y_1,\dots,y_d)  \phi_m(y_1,\dots,y_d)}{\epsilon_m} \phi_m(x_1,\dots,x_d) \nonumber \\
&=&  \frac{E_0^{(1)}}{\sqrt{V_\Omega}} \int_{\Omega_d} d^dy \ G^{(0)}(x_1,\dots,x_d; y_1,\dots,y_d)
\Sigma(y_1,\dots,y_d)   \nonumber 
\eeq

To second order one obtains the equation
\beq
(-\Delta) \Psi_0^{(2)}(x_1,\dots,x_d) + \Psi_0^{(1)}(x_1,\dots,x_d)  = 
\Sigma(x_1,\dots,x_d)  \left( E_0^{(2)} \Psi_0^{(0)}(x_1,\dots,x_d) + E_0^{(1)} \Psi_0^{(1)}(x_1,\dots,x_d)\right)  \nonumber 
\eeq
 
In this case we have
\beq
E_0^{(2)} &=& - E_0^{(1)} \sqrt{V_\Omega}  \frac{\int_\Omega d^dx \ \Sigma(x_1,\dots,x_d) \Psi_0^{(1)}(x_1,\dots,x_d)}{\int_\Omega d^dx \Sigma(x_1,\dots,x_d)} \nonumber  \\
&=& - \frac{\left(E_0^{(1)} \right)^2}{\int_\Omega d^dx \ \Sigma(x_1,\dots,x_d)} \sum_m^\prime
  \frac{\left(\int_\Omega d^dx \Sigma(x_1,\dots,x_d) \   \phi_m(x_1,\dots,x_d)\right)^2}{\epsilon_m}  \nonumber \\
&=&  - \left(E_0^{(1)} \right)^2  \frac{\int_\Omega d^dx \ \int_\Omega d^dy \ \Sigma(x_1,\dots,x_d) \ G^{(0)}(x_1,\dots,x_d; y_1,\dots,y_d) \ \Sigma(y_1,\dots,y_d) }{\int_\Omega d^dx \ \Sigma(x_1,\dots,x_d)}   \nonumber  \\
\Psi_0^{(2)}(x_1,\dots,x_d) &=& \sum_{m\neq 0}  c_m^{(2)} \phi_m(x_1,\dots,x_d) \nonumber \\
&=& \frac{E_0^{(2)}}{\sqrt{V_\Omega}} \sum_{m\neq 0}  \frac{1}{\epsilon_m} \int_{\Omega_d} d^dy \ \Sigma(y_1,\dots,y_d)  \phi_m(y_1,\dots,y_d) \phi_m(x_1,\dots,x_d)\nonumber \\
&+& E_0^{(1)} \sum_{m\neq 0}  \frac{1}{\epsilon_m} \int_{\Omega_d} d^dy \ \Sigma(y_1,\dots,y_d)  \Psi_0^{(1)}(y_1,\dots,y_d)  \phi_m(y_1,\dots,y_d) \phi_m(x_1,\dots,x_d) \nonumber \\
&-& \frac{E_0^{(1)}}{\sqrt{V_\Omega}}  \sum_{m\neq 0}   \frac{\int_{\Omega_d} d^dy \Sigma(y_1,\dots,y_d)  \phi_m(y_1,\dots,y_d)}{\epsilon_m^2} \phi_m(x_1,\dots,x_d) \nonumber  \\
&=& \frac{E_0^{(2)}}{\sqrt{V_\Omega}}\int_{\Omega_d} d^dy \ G^{(0)}(x_1,\dots,x_d; y_1,\dots,y_d) \ \Sigma(y_1,\dots,y_d) \nonumber \\
&+& E_0^{(1)} \int_{\Omega_d} d^dy \  G^{(0)}(x_1,\dots,x_d; y_1,\dots,y_d)  \Sigma(y_1,\dots,y_d)  \Psi_0^{(1)}(y_1,\dots,y_d)  \nonumber \\
&-& \frac{E_0^{(1)}}{\sqrt{V_\Omega}} \int_{\Omega_d} d^dy \  G^{(1)}(x_1,\dots,x_d; y_1,\dots,y_d)  \ \Sigma(y_1,\dots,y_d)  \nonumber 
\eeq

To third order one obtains the equation
\beq
(-\Delta) \Psi_0^{(3)}(x_1,\dots,x_d) + \Psi_0^{(2)}(x_1,\dots,x_d)  =  \Sigma(x_1,\dots,x_d)  \ \sum_{k=1}^3 E_0^{(k)} \Psi_0^{(3-k)}(x_1,\dots,x_d)  \nonumber 
\eeq

In this case we have
\beq
E_0^{(3)} &=& - E_0^{(2)} \sqrt{V_\Omega}  \frac{\int_\Omega d^dx \ \Sigma(x_1,\dots,x_d) \Psi_0^{(1)}(x_1,\dots,x_d)}{\int_\Omega d^dx \Sigma(x_1,\dots,x_d)} \nonumber  \\
&-&   E_0^{(1)} \sqrt{V_\Omega}  \frac{\int_\Omega d^dx \ \Sigma(x_1,\dots,x_d) \Psi_0^{(2)}(x_1,\dots,x_d)}{\int_\Omega d^dx \Sigma(x_1,\dots,x_d)} \nonumber  \\
&=& - \frac{2 E_0^{(2)}  E_0^{(1)}}{\int_\Omega d^dx \Sigma(x_1,\dots,x_d)}  \sum_m^\prime 
  \frac{\left(\int_\Omega d^dx \Sigma(x_1,\dots,x_d) \   \phi_m(x_1,\dots,x_d)\right)^2}{\epsilon_m} \nonumber \\
&-& \frac{\left(E_0^{(1)}\right)^3  }{\int_\Omega d^dx \Sigma(x_1,\dots,x_d)}  \sum_n^\prime \sum_m^\prime
  \frac{\left(\int_\Omega d^dx \Sigma(x) \   \phi_m(x)\right) 
  \left(\int_\Omega d^dx \phi_n(x) \Sigma(x) \   \phi_m(x)\right)
  \left(\int_\Omega d^dx \Sigma(x) \   \phi_n(x)\right)}{\epsilon_m \epsilon_n}  \nonumber\\
&+& \frac{\left(E_0^{(1)}\right)^2  }{\int_\Omega d^dx \Sigma(x_1,\dots,x_d)}  \sum_m^\prime
  \frac{\left(\int_\Omega d^dx \Sigma(x) \   \phi_m(x)\right)^2 }{\epsilon_m^2}  \nonumber
\eeq

Notice that the expressions above can be conveniently expressed in terms of the matrix elements
\beq
\langle m | \Sigma | n\rangle \equiv \int_\Omega d^dx \phi_n(x) \Sigma(x) \   \phi_m(x)  \nonumber 
\eeq
where
\beq
\langle 0 | \Sigma | n\rangle \equiv \frac{1}{\sqrt{V_\Omega}}\int_\Omega d^dx \Sigma(x) \   \phi_m(x)  \nonumber 
\eeq

\beq
E_0^{(2)} &=& - \frac{V_\Omega\left(E_0^{(1)} \right)^2}{\int_\Omega d^dx \ \Sigma(x_1,\dots,x_d)} \sum_m^\prime
  \frac{\langle 0 | \Sigma | m\rangle \langle m | \Sigma | 0\rangle}{\epsilon_m}   \nonumber \\
E_0^{(3)} &=& - \frac{2 V_\Omega E_0^{(2)}  E_0^{(1)}}{\int_\Omega d^dx \Sigma(x_1,\dots,x_d)}  \sum_m^\prime 
  \frac{\langle 0 | \Sigma | m\rangle \langle m | \Sigma | 0\rangle}{\epsilon_m} \nonumber \\
&-& \frac{V_\Omega \left(E_0^{(1)}\right)^3  }{\int_\Omega d^dx \Sigma(x_1,\dots,x_d)}  \sum_n^\prime \sum_m^\prime
  \frac{\langle 0 | \Sigma | m\rangle  \langle m | \Sigma | n\rangle \langle n | \Sigma | 0\rangle}{\epsilon_m \epsilon_n}  \nonumber\\
&+& \frac{V_\Omega \left(E_0^{(1)}\right)^2  }{\int_\Omega d^dx \Sigma(x_1,\dots,x_d)}  \sum_m^\prime
  \frac{\langle 0 | \Sigma | m\rangle \langle m | \Sigma | 0\rangle}{\epsilon_m^2}  \nonumber
\eeq

\section*{Acknowledgments}
This research was supported by the Sistema Nacional de Investigadores (M\'exico).


\begin{thebibliography}{}
\bibitem{Amore13a} P. Amore, "Exact sum rules for inhomogeneous strings", in press, Annals of Physics (2013)
\bibitem{Amore13b} P. Amore, Annals of Physics {\bf 336}, 223-244 (2013)
\bibitem{Gottlieb89} H. P. W.  Gottlieb, Journal of Sound and Vibration {\bf 135}, 79-83 (1989)
\bibitem{Merzbacher70} E. Merzbacher, Quantum mechanics. Second Ed. New York. Wiley. pag.429-430

\end{thebibliography}
\end{document}